\documentclass[conference,final]{IEEEtran}

\ifCLASSINFOpdf
\else
\fi

\usepackage[cmex10]{amsmath}
\usepackage{amsfonts}
\usepackage{amssymb}
\usepackage{graphicx}

\newtheorem{theorem}{Theorem}

\newtheorem{definition}[theorem]{Definition}
\newtheorem{example}[theorem]{Example}
\newtheorem{lemma}[theorem]{Lemma}

\newenvironment{proof}[1][Proof]{\textbf{#1.} }{\ \rule{0.5em}{0.5em}}

\newcommand{\snr}{{P/N}}

\begin{document}

\author{
\IEEEauthorblockN{Hassan Ghozlan}
\IEEEauthorblockA{Department of Electrical Engineering\\
University of Southern California\\
Los Angeles, CA 90089 USA\\
ghozlan@usc.edu}
\and
\IEEEauthorblockN{Gerhard Kramer}
\IEEEauthorblockA{Department of Electrical Engineering\\
University of Southern California\\
Los Angeles, CA 90089 USA\\
gkramer@usc.edu}
}

\IEEEoverridecommandlockouts
\IEEEpeerreviewmaketitle

\title{Interference Focusing for Mitigating Cross-Phase \\ Modulation in a Simplified Optical Fiber Model}

\maketitle

\begin{abstract}
A memoryless interference network model is introduced that is based on non-linear phenomena observed when transmitting information over optical fiber using wavelength-division multiplexing. The main characteristic of the model is that amplitude variations on one carrier wave are converted to  phase variations on another carrier wave, i.e., the carriers interfere with each other through amplitude-to-phase conversion. For the case of two carriers, a new technique called \textit{interference focusing} is proposed where each carrier achieves the capacity pre-log 1, thereby doubling the pre-log of 1/2 achieved by using conventional methods. The technique requires neither channel time variations nor global channel state information. Generalizations to more than two carriers are outlined. 
\end{abstract}

\section{Introduction}
\label{sec:intro}
The additive white Gaussian noise (AWGN) channel, suitably modified, is a good model for many problems encountered in practice. For example, a parallel AWGN channel is accurate for communication over copper cables with orthogonal frequency-division multiplexing (OFDM). An AWGN channel with multiplicative noise models multi-path fading for wireless communication. The capacities \cite{Shannon1948} of these channels have been studied in great detail. In contrast, the capacities of fiber-optic channels have attracted less interest in the information theory community
(see, e.g., the tutorial paper \cite{JLT2010}).
Perhaps this is because it was not until recently that it became necessary to communicate efficiently over fiber; optical fiber has long been viewed as having ``bandwidth to burn". However, the relentless increase in traffic demand and advances in optical technology have made determining fiber capacity of great interest. 

\section{Fiber Channel Models} \label{sec:fiber_channel}
The fiber channel suffers impairments such as propagation loss, dispersion, and Kerr non-linearity. Optical amplifiers such as Erbium-doped fiber amplifiers (EDFAs) compensate the attenuation in fiber links without electronic regeneration, and as a result amplified spontaneous emission (ASE) noise becomes a significant problem. Dispersion arises because the propagating medium absorbs energy through the oscillations of bound electrons, causing a {\it frequency} dependence of the material refractive index~\cite[p.~7]{Agrawal}. The Kerr effect is caused by anharmonic motion of bound electrons in the presence of an intense electromagnetic field, causing an {\it intensity} dependence of the material refractive index~\cite[p.~17, 165]{Agrawal}.

Let $A(z,t)$ be a complex number representing the slowly varying component (or envelope) of a single mode, linearly polarized, electric field at position $z$ and time $t$. Suppose we use a retarded-time reference frame with $T=t-\beta_1 z$ where $\beta_1$ is the reciprocal of the group velocity. Suppose further that the ASE noise is negligible. The evolution of $A(z,T)$ is then governed by the generalized non-linear Schr\"{o}dinger (NLS) equation \cite[p. 44, 50]{Agrawal}:
\begin{align}
	i \frac{\partial A}{\partial z} + \frac{i\alpha}{2} A - \frac{\beta_2}{2} \frac{\partial^2 A}{\partial T^2} 
	+ \gamma |A|^2 A = 0
	\label{eq:nls_eq}
\end{align}
where $i = \sqrt{-1}$, $\alpha$ is the attenuation constant, $\beta_2$ is the group velocity dispersion (GVD) coefficient, $\gamma  = n_2 \omega_0 / (c A_{\text{eff}})$, $n_2$ is the non-linear refractive index, $\omega_0$ is the carrier frequency, $c$ is the speed of light, and $A_{\text{eff}}$ is the effective cross-section area of the fiber. One usually normalizes $A(z,T)$ using $e^{-\alpha z/2}$ which effectively lets one set $\alpha=0$ (see~\cite[p.~50, 64]{Agrawal}).

We are interested in studying the impact of non-linearities, so we consider the {\it simplified} model where $\beta_2 = 0$, i.e., there is no dispersion or completely-compensated dispersion. Eq.~(\ref{eq:nls_eq}) with $\alpha=0$ and $\beta_2 = 0$ has the exact solution \cite[p. 98]{Agrawal}
\begin{align}  \label{eq:solution}
	A(L,T) = A(0,T) e^{i \gamma L |A(0,T)|^2}.
\end{align}
where $L$ is the fiber length.
In other words, Kerr non-linearity leaves the pulse shape unchanged but causes an intensity-dependent phase shift. The phase shift phenomenon is called self-phase modulation (SPM).


Suppose now that two optical fields at different carrier frequencies $\omega_1$ and $\omega_2$ are launched at the same location and propagate simultaneously inside the fiber. The fields interact with each other through the Kerr effect \cite[Ch. 7]{Agrawal}.
Specifically, neglecting fiber losses by setting $\alpha=0$, the propagation is governed by the coupled NLS equations \cite[p. 264, 274]{Agrawal}:
\begin{align}
	& i \frac{\partial A_1}{\partial z}  - \frac{\beta_{21}}{2} \frac{\partial^2 A_1}{\partial T^2} 
	+ \gamma_1 (|A_1|^2 + 2 |A_2|^2) A_1 = 0
	\label{eq:nls2_eq1} \\
	& i \frac{\partial A_2}{\partial z} - \frac{\beta_{22}}{2} \frac{\partial^2 A_2}{\partial T^2} 
	+ \gamma_2 (|A_2|^2 + 2 |A_1|^2) A_2 + i d \frac{\partial A_2}{\partial T} = 0
	\label{eq:nls2_eq2}
\end{align}
where 
$A_j(z,T)$ is the time-retarded, slowly varying component of field $j$, $j=1,2$, the $\beta_{2j}$ are GVD coefficients, the $\gamma_j$ are nonlinear parameters, and $d=\beta_{12}-\beta_{11}$ where the $\beta_{1j}$ are reciprocals of group velocities. We will assume that $\omega_1$ and $\omega_2$ are sufficiently close so that we can set $d=0$. We further simplify and choose $\beta_{21}=\beta_{22}=0$. The coupled NLS equations \eqref{eq:nls2_eq1}-\eqref{eq:nls2_eq2} have the exact solutions \cite[p. 275]{Agrawal}
\begin{align}
	& A_1(L,T) = A_1(0,T) e^{i \gamma_1 L ( |A_1(0,T)|^2 + 2 |A_2(0,T)|^2)}
	\label{eq:xpm_exact1} \\
	& A_2(L,T) = A_2(0,T) e^{i \gamma_2 L ( |A_2(0,T)|^2 + 2 |A_1(0,T)|^2)}
	\label{eq:xpm_exact2}
\end{align}
where $z=0$ is the point at which both fields are launched. Kerr non-linearity again leaves the pulse shapes unchanged but  causes {\it interference} through intensity-dependent phase shifts. The interference phenomenon is called cross-phase modulation (XPM). XPM is an important impairment in optical networks using wavelength-division multiplexing (WDM), see~\cite{JLT2010}.

Equations \eqref{eq:nls2_eq1}-\eqref{eq:xpm_exact2} generalize naturally to launching and receiving fields at different locations, and to using $K$ fields with $K>2$. However, it seems prudent to emphasize that ignoring dispersion, or memory, is considered unrealistic for optical networks. On the other hand, our results do show that a new method called {\it interference focusing} is needed to approach capacity without dispersion. It seems natural to expect that this method will be useful with dispersion also, and this is the subject of ongoing work. The reason we study a {\it memoryless} model is to take a first step in gaining understanding.

To strengthen the link to {\it realistic} channel models, we point out that a 2-parameter phenomenological model that captures the effects of XPM in optically-routed WDM networks was recently proposed in \cite{Goebel}. The model is memoryless in that each received symbol $Y$ is related to the transmitted signal only through the current transmitted symbol $X$ as follows:
\begin{align}
	Y = X e^{i \Phi_{PN}} + Z
	\label{eq:phenom}
\end{align}
where $\Phi_{PN}$ is a Gaussian random variable with variance $c_1 \text{Var}(|X|^2)$ and $c_1$ is a parameter that accounts for system specifications, e.g., the number of WDM channels. $Z$ is AWGN with variance $\sigma^2_Z$ = $N + c_2 \text{Var}(|X|)^3$ where $N$ is the noise variance in the absence of non-linearities and $c_2$ is another system-specific parameter.
The authors of \cite{Goebel} use \eqref{eq:phenom} to accurately predict the channel capacities obtained from full-field numerical simulations reported in \cite{Essiambre2008PRL,Essiambre2009}. We make two observations. First, the WDM channels in \cite{Essiambre2008PRL,Essiambre2009} are made approximately memoryless by using reverse propagation to compensate dispersion. Second, the $\Phi_{PN}$ in \eqref{eq:phenom} is approximately Gaussian if $\Phi_{PN}$ is a sum of many weighted terms of the form $|A_k(0,T)|^2$, $k=1,2,\ldots,K$, similar to \eqref{eq:xpm_exact1}-\eqref{eq:xpm_exact2}.



\section{Interference Network Model}
\label{sec:model}
Equations \eqref{eq:nls2_eq1}-\eqref{eq:xpm_exact2}  and their generalizations to $K$ frequencies motivate the following memoryless interference network model based on sampling the fields $A_k(z,T)$, $k=1,2,\ldots,K$, at $z=0$ and $z=L$. Transmitter $k$ sends a string of symbols $X^n_k = (X_{k,1}, X_{k,2}, \cdots, X_{k,n})$ while receiver $k$ sees $Y^n_k = (Y_{k,1}, Y_{k,2}, \cdots, Y_{k,n})$.
We model the input-output relationship at each time instant $j$ as
\begin{align}
	Y_{k,j} = X_{k,j} \exp\left( i \sum_{\ell=1}^K h_{k\ell} |X_{\ell,j}|^2 \right) + Z_{k,j} 
	\label{eq:channel}
\end{align}
for $k=1,2,\ldots,K$ where $Z_{k,j}$ is circularly-symmetric complex Gaussian noise with variance $N$.
All noise random variables at different receivers and different times are taken to be independent. The terms $\exp(i h_{kk} |X_{k,j}|^2)$ model SPM and the terms $\exp(i h_{k\ell} |X_{\ell,j}|^2)$, $k \neq \ell$, model XPM. We regard the $h_{k\ell}$ as \textit{channel coefficients} that are time invariant. These coefficients are known at the transmitters as well as the receivers, although we shall later see that we need local channel state information only. We use the power constraints
\begin{align} \label{eq:power-constraint}
	\frac{1}{n} \sum_{j=1}^{n} \mathbb{E}\left[ |X_{k,j}|^2 \right] \le P^{(k)}, \quad k=1,2,\ldots,K.
\end{align}

%

\begin{definition}
The \textit{pre-log} $r_k$ achieved by user $k$, whose information rate is $R_k\left(P^{(1)},\ldots,P^{(K)},N\right)$, is
\begin{equation}
r_k = \lim_{P^{(1)}/N,\ldots,P^{(K)}/N \rightarrow \infty} \frac{R_k\left(P^{(1)},\ldots,P^{(K)},N\right)}{\log(P^{(k)}/N)}
\label{eq:prelog_userk_def}
\end{equation}
for $k=1,\ldots,K$.
Thus, a $K$-user transmission scheme may be studied by computing the pre-log $K$-tuple $(r_1,r_2,\ldots,r_K)$.
\end{definition}

One achieves the pre-log $K$-tuple $(1/2,1/2,\ldots,1/2)$ if all users use only amplitude modulation or only phase modulation. The main point of our work is to show that one can, in fact, achieve the ultimate pre-log $K$-tuple $(1,1,\ldots,1)$.

We again emphasize that we have ignored dispersion. Furthermore, the validity of \eqref{eq:channel} depends on the amplification, the network topology, the type of fiber, and so forth. For instance, when performing distributed amplification with stimulated Raman scattering, then additional phase noise should be included at high transmit powers. Also, our analysis assumes $P^{(k)}/N\rightarrow\infty$ for all $k$, and it assumes perfect channel knowledge. Determining the capacity for finite $P^{(k)}/N$ and partial channel knowledge are the subjects of ongoing work.

\section{Two-User Interference Channel}
\label{sec:results}
Consider the 2-user interference channel for which \eqref{eq:channel} without the time indexes becomes
\begin{align}
	Y_1 & = X_1 \exp\left({i h_{11} |X_1|^2+ i h_{12} |X_2|^2}\right) + Z_1 \label{eq:channel2a} \\
	Y_2 & = X_2 \exp\left({i h_{21} |X_1|^2+ i h_{22} |X_2|^2}\right) + Z_2. \label{eq:channel2b}
\end{align}
We propose an \textit{interference focusing} scheme in which the transmitters \textit{focus} their phase interference on one point by constraining their transmitted signals to satisfy
\begin{align}
	h_{21} |X_1|^2 & = 2 m \pi, ~ m=1,2,3,\ldots \label{eq:ring-constraint-1} \\
	h_{12} |X_2|^2 & = 2 \tilde{m} \pi, ~ \tilde{m}=1,2,3,\ldots \label{eq:ring-constraint-2} 
\end{align}
In other words, the transmitters use \textit{multi-ring modulation} with specified spacings between the rings.\footnote{Multi-ring modulation was used in~\cite{JLT2010,Essiambre2008PRL,Essiambre2009} for symmetry and computational reasons only. We here find that it is useful for improving rate.} We thereby remove XPM interference and \eqref{eq:channel2a}-\eqref{eq:channel2b} reduce to
\begin{align}
	Y_k = X_k e^{i h_{kk} |X_k|^2} + Z_k, \quad k=1,2.
\end{align}
This channel is effectively an AWGN channel since $h_{kk}$ is known by receiver $k$ and the SPM phase shift is determined by the desired signal $X_k$.

It remains to show that the pre-log pair $(r_1,r_2)=(1,1)$ is achieved under the constraints \eqref{eq:ring-constraint-1}-\eqref{eq:ring-constraint-2}. We show this in two steps: we first determine the information rate for one ring and then extend the analysis to many rings.

\subsection{One Ring}
\label{sec:onering}
Consider the AWGN channel $Y=X+Z$ with $X = \sqrt{P} e^{i \Phi}$ where $\Phi$ is uniformly distributed on the interval $[0,2\pi)$. The achievable rate $R$ is given by
\begin{align}
	R = I(X;Y) 
	&= h(Y) - h(Y|X) \nonumber \\
	&= \mathbb{E}[-\log p_Y(Y)] - \log(\pi e N)	 \label{eq:information}		
\end{align}
The probability density of $Y$ can be shown to be \cite[p. 688]{JLT2010}
\begin{align}
	p_Y(y) 
	&=  \frac{1}{\pi N} e^{-(y_A^2+P)/N} I_0\left(\frac{2 y_A \sqrt{P} }{N}\right)
\end{align}
where $I_0(\cdot)$ is the modified Bessel function of the first kind of order zero
and $y_A = |y|$.
Therefore, we have
\begin{align} \label{eq:hY}
	h(Y) =  \mathbb{E}\left[ 
	-\log\left(\frac{1}{\pi N} e^{-(Y_A^2+P)/N} I_0\left(\frac{2 Y_A \sqrt{P} }{N}\right) \right)
	\right] .
\end{align}

Next, we derive an upper bound on $I_0(z)$ that we will use  in the process of lower bounding $h(Y)$.
\begin{lemma} \label{lemma:lemma}
We have
\begin{align}
	I_0(z) \leq \frac{\sqrt{\pi}}{2} \frac{e^z}{\sqrt{z}}, \quad z \ge 0.
\end{align}
\end{lemma}
\begin{proof}
We have $\cos x \leq 1 - 4 x^2/\pi^2$ for $0 \leq x \leq \pi/2$ by using the infinite product form for cosine
\begin{align}
	\cos x = \prod_{n=1}^\infty \left[ 1- \frac{4x^2}{\pi^2 (2 n -1)^2}\right].
\end{align}
We thus have
\begin{align}
	I_0(z)
	& = \frac{1}{\pi} \int_{0}^{\pi} e^{z \, cos \, \theta} \, d\theta \nonumber \\
	& \le \frac{1}{\pi} \left[ \int_{0}^{\pi/2} e^{z (1- 4 \theta^2/\pi^2)} \, d\theta + \int_{\pi/2}^{\pi} e^{0} \, d\theta \right] \nonumber \\
	& = \frac{\sqrt{\pi}}{4} \frac{e^z}{\sqrt{z}} \left( 1 - 2 Q(\sqrt{2z}) \right) + \frac{1}{2} \nonumber
\end{align}
where
$Q(z) = \int_{z}^{\infty} \frac{1}{\sqrt{2 \pi}} e^{-x^2/2}  dx$. Finally, we observe that $Q(z) \ge 0$ and $\sqrt{\pi}e^z/(4\sqrt{z}) \ge 1/2$ for $z \geq 0$.
\end{proof}

Using Lemma~\ref{lemma:lemma} in~\eqref{eq:hY}, we have
\begin{align}
	h(Y) 
	&\geq \mathbb{E}\left[ 
	-\log\left(\frac{1}{2 \sqrt{\pi} N} \frac{e^{-(Y_A - \sqrt{P})^2 /N}}{\sqrt{2 Y_A \sqrt{P} /N}} \right)
	\right] \\
	& \geq
	\mathbb{E}\left[  \log\left(2 \sqrt{\pi} N {\sqrt{2 Y_A \sqrt{P}/N}} \right) \right] \\
	& =
	\frac{1}{4} \log\left(32 \pi^2 P N^3 \right)
	+
	\frac{1}{2} \mathbb{E} \left[ \log\left(Y_A \sqrt{\frac{2}{N}}\right) \right].
	\label{eq:entropy_y_lowerbound}
\end{align}
The last expectation in \eqref{eq:entropy_y_lowerbound} is
\begin{align}
	\int_{y_A=0}^{\infty} \frac{2 y_A}{N} e^{-(y_A^2+P)/N} 
	I_0\left(\frac{2 y_A \sqrt{P}}{N}\right) 
	\log\left(y_A \sqrt{\frac{2}{N}}\right) \, dy_A.
	\label{eq:elogya}
\end{align}
Setting $z = y_A \sqrt{2/N}$ and $\nu^2=2P/N$, expression (\ref{eq:elogya}) is
\begin{align}
	& \frac{\log(e)}{2} \int_{z=0}^{\infty} 2 z \exp\left(-\frac{z^2 + \nu^2}{2}\right) 
	I_0\left( z \nu \right) \ln(z) \, dz \nonumber \\
	& \quad = \frac{\log(e)}{2} \left[ \Gamma\left(0,\frac{P}{N}\right) + \ln \left(\frac{2P}{N}\right) \right]
	\label{eq:igamma}
\end{align}
where $\Gamma(a,x)$ is the upper incomplete Gamma function \cite[p. 260]{Abramowitz1972} and where the second step follows by \cite{Mathematica7}.
%
Inserting \eqref{eq:igamma} into (\ref{eq:entropy_y_lowerbound}), and then (\ref{eq:entropy_y_lowerbound}) into \eqref{eq:information}, gives
\begin{align}
  I(X;Y) & \ge \frac{1}{2} \log\left(\frac{8}{\pi e^2} \frac{P}{N} \right)
	          + \frac{\log(e)}{4} \, \Gamma\left(0,\frac{P}{N}\right) \\
	& \ge \frac{1}{2} \log\left(\frac{8}{\pi e^2} \frac{P}{N} \right)  
	\label{eq:one_ring_lb}
\end{align}
where we have made use of $\Gamma\left(0,x\right) \ge 0$ for $x\ge0$.\footnote{Note that $\lim_{x \rightarrow \infty} \Gamma\left(0,x\right) = 0$.} The desired pre-log therefore satisfies 
\begin{align}
  r = \lim_{\snr \rightarrow \infty} \frac{R(\snr)}{\log(\snr)} \geq \frac{1}{2}.
\end{align}

\subsection{Multiple Rings}
\label{sec:multiring}
Consider multiple rings with $X = \sqrt{P_j} e^{i \Phi}$, $j=1,\ldots,J$, where $J$ is the number of rings. The power levels $P_j$ allowed under interference focusing take the form $m p_0$ where $m$ is a positive integer and $p_0$ is the minimum (non-zero) power level that depends on the channel coefficients. For example, for the 2-user interference channel $p_0 = 2 \pi / h_{21}$ for transmitter 1 and $p_0 = 2 \pi / h_{12}$ for transmitter 2. The power levels must further satisfy $\mathbb{E}[|X|^2]\le P$ where $P$ is the power constraint of the user being considered (our pre-log analysis is based on a point-to-point AWGN channel because interference is removed by interference focusing).
The achievable rate $R$ is given by
\begin{align}
	R = I(X;Y) 
	&= I(X_A , \Phi;Y) \nonumber \\
	& = I(X_A;Y) + I(\Phi;Y|X_A) \label{eq:mult-ring-rate}
\end{align}
where $X=X_A e^{i \Phi}$. The term $I(X_A ; Y)$ can be viewed as the amplitude contribution while the term $I(\Phi ; Y | X_A)$ is the phase contribution.

Suppose that, for simplicity, we choose the rings to be spaced uniformly in amplitude as
\begin{align} \label{eq:power-levels}
	P_j = a j^2 \, p_0
\end{align}
where $a$ is a positive integer. We further use a uniform frequency of occupation of rings with $P_{X_A}(\sqrt{P_j}) = 1/J$, $j=1,2,\ldots,J$. The power constraint is therefore
\begin{align} \label{eq:pwr_constraint_uniform}
	\frac{1}{J} \sum_{j=1}^J a j^2 \, p_0 \le P .
\end{align}

\subsubsection{Phase Contribution}
Using \eqref{eq:one_ring_lb}, we have
\begin{align}
	I(\Phi_A;Y|X_A) 
	& = \sum_{j=1}^J \frac{1}{J} \, I(\Phi_A;Y|X_A=\sqrt{P_j}) \\
	& \geq \sum_{j=1}^J \frac{1}{J}  \, \frac{1}{2} \log\left( \frac{8}{\pi e^2} \frac{P_j}{N} \right).
	\label{eq:sum-ln}
\end{align}
We show in the Appendix that by choosing $a$ in \eqref{eq:power-levels} to scale as $N\log(\snr)$, and choosing $J$ to satisfy \eqref{eq:pwr_constraint_uniform}, then we have
\begin{align}  \label{eq:phase-prelog}
	\lim_{\snr \rightarrow \infty}  \frac{\frac{1}{2} \frac{1}{J} \sum_{j=1}^J \log(P_j/N)}{\log(\snr)} \ge \frac{1}{2}
\end{align}
The pre-log of the phase contribution is therefore at least $1/2$.

\subsubsection{Amplitude Contribution}
We have
\begin{align}
	I(X_A;Y)  = H(X_A) - H(X_A|Y)
\end{align}
where $H(X_A) = \log(J)$. We show in the Appendix that $J$ scales as $\sqrt{(\snr)/\log(\snr)}$ if $a$ scales as $N\log(\snr)$. We bound $H(X_A|Y)$ using Fano's inequality as
\begin{align}
	H(X_A|Y) &\le H(X_A|\hat{X}_A) \nonumber \\
	&\le H(P_e) + P_e \log(J-1)
	\label{eq:fano}	
\end{align}
where $\hat{X}_A$ is any estimate of $X_A$ given $Y$, $P_e=\Pr[\hat{X}_A \ne X_A]$ and $H(P_e)$ is the binary entropy function with a general logarithm base. Suppose we use the minimum distance estimator
\begin{align}
	\hat{X}_A = \arg \min_{x_A \in \mathcal{X}_A} |Y_A-x_A|
	\label{eq:min_distance_dec}
\end{align}
where $Y_A=|Y|$ and $\mathcal{X}_A = \{\sqrt{P_j}:j=1,\ldots,J\}$. We show in the Appendix that
\begin{align} \label{eq:Pe-bound}
	P_e & \le \frac{2}{J} \sum_{j=2}^{J} \exp\left(-\frac{\Delta_j^2}{4}\right)	
\end{align}
where $\Delta_j = (\sqrt{P_j} - \sqrt{P_{j-1}})/\sqrt{N}$. 
For the power levels (\ref{eq:power-levels}), we have $\Delta_j = \sqrt{a p_0/N}$ for all $j$, and hence
\begin{align} \label{eq:Pe-bound-2}
	P_e 
	\le \frac{2(J-1)}{J} \exp\left(-\frac{a p_0}{4 N}\right) .
\end{align}
We see from \eqref{eq:Pe-bound-2} that $\lim_{\snr \rightarrow \infty} P_e = 0$ if $a$ scales as $N\log(\snr)$
(recall that $J$ scales as $\sqrt{(\snr)/\log(\snr)}\,$). 
We thus have $\lim_{\snr \rightarrow \infty} H(X_A|Y) = 0$ by using (\ref{eq:fano}).
Consequently, we have
\begin{align} \label{eq:amplitude-prelog}
	\lim_{\snr \rightarrow \infty} \frac{I(X_A;Y)}{\log(\snr)} 
	= \lim_{\snr \rightarrow \infty} \frac{\log(J)}{\log(\snr)} 
	= \frac{1}{2}.
\end{align}

We conclude from \eqref{eq:mult-ring-rate}, \eqref{eq:phase-prelog}, and \eqref{eq:amplitude-prelog} that interference focusing achieves the largest-possible pre-log of 1. Each user can therefore exploit all the phase and amplitude degrees of freedom simultaneously.

\section{$K$-User Interference Network}
\label{sec:discussion}
We outline how to apply interference focusing to problems with $K>2$. Define the interference phase vector
\begin{align}
   \underline{\Psi} \stackrel{\Delta}{=} [\Psi_1,\Psi_2,\ldots,\Psi_K]^T
\end{align}
where $\Psi_k = \sum_{\ell=1}^K h_{k\ell} |X_{\ell}|^2$ and the instantaneous power vector
\begin{align}
  \underline{\Pi} \stackrel{\Delta}{=} \left[ |X_1|^2,\ldots,|X_K|^2 \right]^T.
\end{align}
The relation between the $\underline{\Psi}$ and $\underline{\Pi}$ in matrix form is
\begin{align}
	\underline{\Psi} = H_{SP} \, \underline{\Pi}  + H_{XP} \, \underline{\Pi} 
\end{align}
where $H_{SP}$ is a diagonal matrix that accounts for SPM and $H_{XP}$ is a zero-diagonal matrix that accounts for XPM.

\begin{example} \label{exampe:HXP}
Suppose the XPM matrix for a 3-user interference network is
\begin{align}
	H_{XP} = \left[
	\begin{array}{ccc}
	0       &      1/2     &      3/5     \\
	3/4     &      0       &      2/3     \\
	5/6     &      1/5     &      0   	
	\end{array}
	\right]
\end{align}
Suppose that each transmitter knows the channel coefficients between itself and all the receiving nodes. The transmitters can thus use power levels of the form
\begin{align}
   \underline{\Pi} & = 2 \pi \cdot \left[ \, \text{lcm}(4,6) m_1, \text{lcm}(2,5) m_2, \text{lcm}(5,3) m_3 \,\right] \nonumber \\
   & = 2 \pi \cdot \left[\,  12 m_1, 10 m_2, 15 m_3 \,\right] 
\end{align}
where $\text{lcm}(a,b)$ is the least common multiple of $a$ and $b$, and $m_1,m_2,m_3$ are positive integers. We thus have
\begin{align}
	H_{XP} \, \underline{\Pi} 
	=
	2 \pi
	\left[
	\begin{array}{ccc}
	0      &      5     &      9     \\
	9   	 &      0     &     10     \\
	10     &      2     &      0   	
	\end{array}
	\right]
	\left[
	\begin{array}{c}
	m_1	\\
	m_2	\\
	m_3     
	\end{array}
	\right]
\end{align}
which implies that the phase interference has been eliminated.
\end{example}

Example \ref{exampe:HXP} combined with an analysis similar to Section \ref{sec:results} shows that interference focusing will give each user a pre-log of $1$ even for $K$-user interference networks. However, the XPM coefficients $h_{k\ell}$ must be \textit{rationals}. Modifying interference focusing for \textit{real-valued} XPM coefficients is an interesting problem. It is clear from Example \ref{exampe:HXP} that interference focusing does not require global channel state information.

\section{Conclusion}
\label{sec:conclusion}
We introduced an interference network model based on a simplified optical fiber model. We assumed that there was no dispersion, or that dispersion was compensated. The non-linear nature of the fiber-optic medium causes the users to suffer from amplitude-dependent phase interference. We introduced a new technique called interference focusing that lets the users take full advantage of all the available amplitude and phase degrees of freedom. Several generalizations are interesting to study further, e.g., introduce group velocity, focus interference on multiple points, study low and intermediate signal-to-noise ratio, investigate partial channel knowledge, and so on.

\section*{Acknowledgment}
\label{sec:ack}
H. Ghozlan was supported by a USC Annenberg Fellowship. G. Kramer was supported by NSF Grant CCF-09-05235. We are grateful to  the reviewers for providing constructive criticisms that helped to improve the paper.


\appendix
\section*{Phase Modulation with Multiple Rings}
We derive the key relations to prove that uniform phase modulation contributes 1/2 to the pre-log for specially-chosen multi-ring modulations. For (\ref{eq:pwr_constraint_uniform}) we compute
\begin{align}
	\frac{1}{J} \sum_{j=1}^{J} a p_0 \, j^2 = a p_0 \frac{(J+1)(2J+1)}{6}
\end{align}
so to satisfy the power constraint we choose\footnote{The solution for $J$ should be positive and rounded down to the nearest integer but we ignore these issues for notational simplicity.}
\begin{align}
	J = \frac{-3  + \sqrt{1 + 48 P/(a p_0)}}{4}.
\end{align}
We choose $a$ to scale as $N\log(\snr)$ so $J$ scales as $\sqrt{(\snr)/\log(\snr)}$.
%
%
Next, consider the sum in \eqref{eq:phase-prelog}. The logarithm is an increasing function so we have
\begin{align}
	\sum_{j=1}^J \log\left(\frac{a j^2 p_0}{N}\right)
	&\geq \int_{x=0}^J \log\left(\frac{a x^2 p_0}{N}\right) dx \\
	&= J \left( \ln(a J^2 p_0/N) - 2 \right) \log(e)
\end{align}
We can therefore write
\begin{align}
	\lim_{\frac{P}{N} \rightarrow \infty}  \frac{\frac{1}{2} \frac{1}{J} \sum_{j=1}^J \log(P_j/N)}{\log(\snr)} 
	&\ge \lim_{\frac{P}{N} \rightarrow \infty} \frac{\frac{1}{2} \log(a J^2 p_0/N)}{\log(\snr)} 
	= \frac{1}{2} \label{eq:amp-scaling}
\end{align}
where \eqref{eq:amp-scaling} follows because $a$ scales as $N\log(\snr)$, $J^2$ scales as $(\snr)/\log(\snr)$,  and $p_0$ is independent of $P$ and $N$.

\section*{Minimum Distance Estimator}
We derive the bound \eqref{eq:Pe-bound} for the estimator (\ref{eq:min_distance_dec}). Let $P_{e,j}$ be the error probability when $X_A=\sqrt{P_j}$. We have $P_e=\sum_{j=1}^J \frac{1}{J} P_{e,j}$ and
\begin{align}
	P_{e,j} & = \left\{ \begin{array}{ll}
	\Pr\left(Y_A \ge \frac{\sqrt{P_1}+\sqrt{P_{2}}}{2} \right), & j=1 \\
	 \Pr\left(Y_A \le \frac{\sqrt{P_{K-1}}+\sqrt{P_K}}{2} \right), & j=J \\
	 \Pr\left(Y_A \le \frac{\sqrt{P_{j-1}}+\sqrt{P_j}}{2} \right) & \\
	\quad + \Pr\left(Y_A \ge \frac{\sqrt{P_j}+\sqrt{P_{j+1}}}{2} \right), & \text{otherwise}.
	\end{array} \right.
\end{align}
Conditioned on $X_A=\sqrt{P_j}$, $Y_A$ is a Ricean random variable, and hence we compute \cite[p. 50]{ProakisSalehi2008}
\begin{align} \label{eq:interval-error}
	\Pr\left(Y_A \ge \frac{\sqrt{P_j}+\sqrt{P_{j+1}}}{2} \right)
	= Q\left(\frac{\sqrt{P_j}}{\sqrt{N/2}},\frac{\sqrt{P_j}+\sqrt{P_{j+1}}}{2 \sqrt{{N/2}} }\right)
\end{align}
where $Q(a,b)$ is the Marcum Q-function \cite{Corazza2002}.
Consider the following bounds.
\begin{itemize}
\item Upper bound for $b>a$ \cite[UB1MG]{Corazza2002}
\begin{align}
	Q(a,b) \le \exp\left(-\frac{(b-a)^2}{2}\right) .
	\label{eq:marcum_q_ub}
\end{align}

\item Lower bound for $b<a$ \cite[LB2aS]{Corazza2002}
\begin{align}
	& Q(a,b) \nonumber \\
	& \ge 1 - \frac{1}{2} \left[ \exp\left(-\frac{(a-b)^2}{2}\right) - \exp\left(-\frac{(a+b)^2}{2}\right) \right] .
	\label{eq:marcum_q_lb}
\end{align}
\end{itemize}
The bound (\ref{eq:marcum_q_lb}) implies
\begin{align}
	1 - Q(a,b) \le \exp\left(-\frac{(a-b)^2}{2}\right) .
	\label{eq:marcum_q_lb_loosened}
\end{align}

We use \eqref{eq:interval-error} and \eqref{eq:marcum_q_ub}  to write
\begin{align}
	\Pr\left(Y_A \ge \frac{\sqrt{P_j}+\sqrt{P_{j+1}}}{2} \right)\
	\le \exp\left(-\frac{\Delta_{j+1}^2}{4}\right)	.
\end{align}
where $\Delta_j = (\sqrt{P_j} - \sqrt{P_{j-1}})/\sqrt{N}$. Similarly, we use inequality \eqref{eq:marcum_q_lb_loosened} to write
\begin{align}
	\Pr\left(Y_A \le \frac{\sqrt{P_{j-1}}+\sqrt{P_j}}{2} \right)
	& \le \exp\left(-\frac{\Delta_j^2}{4}\right) .
\end{align}
Collecting our results, we have
\begin{align}
	P_e 
	& \leq \frac{1}{J} \left[ 
	\exp\left(-\frac{\Delta_2^2}{4}\right)	
	+ \sum_{j=2}^{J-1} \exp\left(-\frac{\Delta_j^2}{4}\right)			
	\right. \nonumber \\ & \left.
	\qquad + \sum_{j=2}^{J-1} \exp\left(-\frac{\Delta_{j+1}^2}{4}\right)	
	+ \exp\left(-\frac{\Delta_J^2}{4}\right)
	\right] \nonumber \\
	& = \frac{2}{J} \sum_{j=2}^{J} \exp\left(-\frac{\Delta_j^2}{4}\right).
\end{align}

\bibliographystyle{unsrt}
\bibliography{optic_ref4}

\end{document}